\documentclass{aastex631}

\usepackage{bm}
\usepackage{xcolor}
\usepackage{multibib}
\newcites{main}{References}
\newcites{appendix}{References (Appendix)}
\usepackage{subfigure}
\usepackage{soul}
\usepackage{tcolorbox}
\usepackage{amsmath}
\usepackage{verbatim}
\newcommand{\beq}{\begin{equation}}
\newcommand{\eeq}{\end{equation}}
\newcommand{\bseq}{\begin{subequations}}
\newcommand{\eseq}{\end{subequations}}
\newcommand{\bary}{\begin{eqnarray}}
\newcommand{\eary}{\end{eqnarray}}
\newcommand{\bwt}{\begin{widetext}}
\newcommand{\ewt}{\end{widetext}}

\begin{document}
\title{
Deciphering the $\sim 18$ TeV photons from GRB 221009A}

\author{Sarira Sahu}
\email{sarira@nucleares.unam.mx}
\affiliation{Instituto de Ciencias Nucleares, Universidad Nacional Aut\'onoma de M\'exico, \\
Circuito Exterior, C.U., A. Postal 70-543, 04510 Mexico DF, Mexico}

\author{B. Medina-Carrillo}
\email{benjamin.medina@cinvestav.mx}
\affiliation{Departamento de Física Aplicada,
Centro de Investigación y de Estudios Avanzados del IPN,\\
Unidad Mérida. A.P. 73, Cordemex,
Mérida, Yucatán 97310, México
}

\author{G. Sánchez-Colón}
\email{gabriel.sanchez@cinvestav.mx}
\affiliation{Departamento de Física Aplicada,
Centro de Investigación y de Estudios Avanzados del IPN,\\
Unidad Mérida. A.P. 73, Cordemex,
Mérida, Yucatán 97310, México
}

\author{Subhash Rajpoot}
\email{Subhash.Rajpoot@csulb.edu}
\affiliation{Department of Physics and Astronomy, California State University,\\ 
1250 Bellflower Boulevard, Long Beach, CA 90840, USA
}

\begin{abstract}
On 9 October, 2022, an extremely powerful gamma-ray burst, GRB 221009A, was detected by several instruments. Despite being obstructed by the Milky Way galaxy, its afterglow outburst outshone all other GRBs seen before. LHAASO detected several thousands very high energy photons extending up to 18 TeV. Detection of such energetic photons are unexpected due to the large opacity of the Universe. It is possible that in the afterglow epoch the intrinsic very high energy photon flux from the source might have increased manifolds, which could compensate the attenuation by pair-production with the extragalactic background light. We propose such a scenario and show that very high energy photons can be observed on the Earth from the interaction of very high energy protons with the seed synchrotron photons in the external forward shock region of the GRB jet.
\end{abstract}

\keywords{Particle astrophysics (96), Blazars (164), Gamma-ray bursts (629), Relativistic jets (1390)}

\section{Introduction}

On October 9, 2022, at T0 = 13 : 16 : 59.000 UT~\citep{Circular32636}, a long duration gamma-ray burst (GRB), identified as GRB 221009A (also known as Swift J1913.1+1946) was detected in the direction of the constellation Sagitta by the Gamma-ray Burst Monitor (GBM)~\citep{Meegan_2009} onboard the Fermi Gamma-ray Space Telescope. The prompt emission was also detected by several other space observatories, such as the Fermi Large Area Telescope (LAT), Swift~\citep{Circular32632, Circular32688}, AGILE~\citep{Circular32650, Circular32657}, INTEGRAL~\citep{Circular32660}, Solar Orbiter~\citep{Circular32661}, SRG~\citep{Circular32663}, Konus~\citep{Circular32668}, GRBAlpha~\citep{Circular32685}, and STPSat-6~\citep{Circular32746}. GRB 221009A is located at the coordinate RA = 288.282 and Dec = 19.495~\citep{Circular32658}. The Fermi-LAT detected the most energetic photon of energy 99.3 GeV (at $t_0+240$ s). It is the highest energy photon ever detected by Fermi-LAT from a GRB in the prompt phase~\citep{Circular32637, Circular32658}. The afterglow emission was also observed at different wavelengths~\citep{Das:2022gon}, and the optical follow-up observation estimated a very low redshift of $z\simeq 0.151$~\citep{Circular32648}. The total emitted isotropic-equivalent gamma-ray energy from GRB 221009A is estimated to be $(2-6)\times 10^{54}$ erg~\citep{Circular32648, Circular32762}. This is the brightest, long-duration GRB, and arguably, one of the nearest, and possibly, the most energetic GRB ever observed. It has also been reported that GRB 221009A produced a significant ionization of the Earth's lower ionosphere ($\sim 60-100$ km)~\citep{Hayes_2022} and is the strongest ionization effect ever recorded from a GRB.


The Large High Altitude Air Shower Observatory (LHAASO) with the water Cherenkov detector array (WCDA) and the larger air shower kilometer square area (KM2A) detector observed more than 5000 very high energy (VHE) photons within $T_0+2000$ s in the 500 GeV to 18 TeV energy range, making them the most energetic photons ever observed from a GRB~\citep{Circular32677}. Surprisingly, the ground-based Cherenkov detector Carpet-2 at Baksan Neutrino Observatory reported the detection of undoubtedly a very rare air-shower originating from a  251 TeV photon 4536 s after the GBM trigger from the direction of the GRB 221009A~\citep{2022ATel15669....1D}. Observations of these unusually  VHE gamma-rays by LHAASO and Carpet-2 from GRB 221009A are incomprehensible, and led to the speculations of non-standard physics explanations of these observed events. However, there is a caveat concerning the observation of 251 TeV gamma-ray. The angular resolution of the Carpet-2 is several degrees and the two previously reported Galactic VHE sources, 3HWC J1928+178 and LHASSO J1929+1745, are located close to the position of the GRB 221009A~\citep{2022ATel15675....1F}. It remains uncertain whether the observed 251 TeV photon is from the GRB 221009A or from either of these Galactic sources. Nevertheless, the temporal and spatial coincidence of this event with the GRB 221009A is worth exploring~\citep{2022arXiv221011261F,AlvesBatista:2022kpg,2022arXiv221014243M}. In the present context, we will delve into the VHE emission observed by LHAASO.

The VHE $\gamma$-rays observed by the Cherenkov telescopes from the extragalactic sources undergo energy-dependent attenuation by interacting with the extragalactic background light (EBL) through electron–positron pair production~\citep{1992ApJ...390L..49S, 2012Sci...338.1190A}. As a result, the shape of the spectrum at very high energies changes significantly. Several well known EBL models have been developed to study the attenuation at different redshifts.These models have been used successfully by the the highly sensitive Imaging Atmospheric Cherenkov Telescopes (IACTs) such as VERITAS~\citep{Holder:2008ux}, HESS~\citep{Hinton:2004eu}, and MAGIC~\citep{Cortina:2004qt}, to analyze the observed VHE gamma-rays from sources of different redshifts. The observed VHE gamma-ray flux from the source can be written in terms of the intrinsic flux $F_{in}$ and the survival probability of the VHE photon as~\citep{Hauser:2001xs}
\beq
F_{\gamma}(E_{\gamma})=F_{in}(E_{\gamma})\, e^{-\tau_{\gamma\gamma}(E_{\gamma})},
\label{obsflux}
\eeq
where $E_{\gamma}$ is the observed VHE photon energy and $\tau_{\gamma\gamma}$ is the optical depth for the pair-production process. The optical depth for a 18 TeV photon at a redshift of $z=0.151$ is 18.3 in EBL model of~\cite{Franceschini:2008tp} and 19.4 in the EBL model of~\cite{Dominguez:2010bv} which corresponds to the survival probability of the VHE photon $e^{-\tau_{\gamma\gamma}} \sim 1.1\times 10^{-8}$ and $3.6\times 10^{-9}$ respectively in both these models. Thus, for a 18 TeV photon energy, the observed flux will be suppressed by a factor of $\sim 10^{-9}-10^{-8}$. As the observation of 18 TeV photon from a source at redshift $z=0.151$ is difficult to comprehend, it is viewed as signature of new physics such as Lorentz invariance violation~\citep{2022arXiv221011376Z,2022arXiv221006338L,2022arXiv221007172B,2022arXiv221011261F}, oscillation of photon to a pseudo-scalar particle (axion-like particle (ALP))~\citep{2022arXiv221005659G,2022arXiv221008841L,2022arXiv221009250T}, ALP abundance  enhanced with its mass caused by a first-order phase transition in a hidden sector~\citep{Nakagawa:2022wwm}, heavy neutrino as the means of propagation to avoid the energy attenuation~\citep{Cheung:2022luv}, and sterile neutrinos produced via mixing with active neutrinos~\citep{Brdar:2022rhc}. On the other hand, from the standard physics point of view, these gamma-rays are argued to be the secondaries arising from the interactions between the ultra-high energy cosmic rays emanating from GRB 221009A and the cosmological photon background on their way to the Earth~\citep{AlvesBatista:2022kpg}. Also, observation of neutrinos from such a bright GRB is discussed~\citep{2022arXiv221015625M}. 


Since the VHE spectra of most of the flaring high energy blazars (HBLs) of different redshifts are explained very well using the EBL models of~\cite{Franceschini:2008tp} and~\cite{Dominguez:2010bv}, 
the recent observation of $\sim 18$ TeV photon from the GRB 221009A falls short of this expectation. 
The obvious question is, can it be due to the intrinsic flux from the source? If we look into Eq.(\ref{obsflux}), the depletion in the flux due to $e^{-\tau_{\gamma\gamma}}$ can, in principle, be compensated by increasing the intrinsic flux. However, this may not be possible in most of the situations. As noted previously, GRB 221009A is very special as its afterglow outburst outshone all other GRBs seen before, despite the fact that GRB 221009A is obstructed by the Milky Way galaxy. Furthermore, the burst was so powerful that it ionized Earth's atmosphere and disrupted long wave radio communications. It is estimated that, at low redshifts, such energetic GRBs are extremely rare events and may occur once in a century~\citep{Circular32793}. Thus, it is possible that the intrinsic VHE flux from the source might have increased manifolds, which could compensate the depletion from the EBL effect. In this letter, we would like to pursue such a scenario and its impact on the observation of $\sim 18$ TeV photons by LHAASO.

\section{Common features of Blazar and GRB}

The emission mechanisms in blazars (a subclass of active galactic nuclei (AGN)) and GRBs have many features in common~\citep{Urry_1995, Gehrels:2013xd}. Such common features are found to prevail in the synchrotron luminosity and Doppler factor between GRBs and active galactic nuclei (AGNs)~\citep{Wu_2011}. In several studies it was observed that the jets in blazars and GRBs share common features despite large differences in their masses and bulk Lorentz factors~\citep{2012Sci...338.1445N, 2011ApJ...726L...4W, Wu:2015opa}. It is  natural to use such mechanisms and processes to study the multi-TeV flaring of high energy blazars to study the afterglow phases of GRBs.

Previously we have used the photohadronic process to study the multi-TeV flaring from HBLs~\citep{Sahu:2019lwj, Sahu_2019,10.1093/mnras/staa023}. In the photohadronic scenario, protons in the blazar jet are accelerated to very high energies and then collide with the background seed photons to produce $\Delta$-resonance ($p\gamma\rightarrow\Delta^+$) with the following kinematical condition~\citep{Sahu:2019lwj}
\beq
E_p \epsilon_{\gamma} = 0.32\, \Gamma\,{\cal D}(1+z)^{-2} \, \mathrm{{GeV}^2},
\label{KinemCond}
\eeq
where $E_p$ and $\epsilon_{\gamma}$ are the proton energy and the background seed photon energy respectively in the observer's frame. In the process, the observed VHE photon carries about 10\% of the proton energy, $E_{\gamma}\simeq 0.1 Ep$. The bulk Lorentz factor and the Doppler factor respectively are given by $\Gamma$ and ${\cal D}$. As the jets of the observed HBLs and GRBs beam towards us, $\Gamma\simeq{\cal D}$. The $\Delta$-resonances decay to neutral pions that subsequently decay to VHE gamma-rays. These are the blueshifted photons observed by the Cherenkov telescopes on Earth. This model is very successful in explaining the VHE gamma-ray spectra from several HBLs, and the intrinsic flux $F_{\mathrm{in}}$ is given by
\beq
F_{in}=F_0\, E^{-\delta+3}_{\gamma,TeV},
\label{eq:fluxgeneral}
\eeq
where $E_{\gamma,TeV}$ is the photon energy in TeVs. The normalization constant $F_0$ can be fixed from the observed spectrum and the spectral index $\delta=\alpha+\beta$ is the free parameter in the model~\citep{Sahu:2019lwj, Sahu_2019}. Note that $F_{in}$ is independent of $\Gamma$ and $\cal D$. The high energy protons in the jet have a power-law differential spectrum $dN/dE_p\propto E^{-\alpha}_p$, $E_p$ is the proton energy and we take $\alpha=2$~\citep{1993ApJ...416..458D}, a generally accepted value. For HBLs, the seed photon flux also follows a power-law $\Phi_{\gamma}\propto \epsilon^{\beta}_{\gamma}\propto E^{-\beta}_{\gamma}$~\citep{Sahu:2019lwj, Sahu_2019}. For HBLs, the value of $\delta$ always lies in the range $2.5\le \delta \le 3.0$, which corresponds to a $\beta$ value in the range $0.5\le \beta \le 1.0$, indicating that the seed photons are in the low energy tail region of the SSC spectrum~\citep{Sahu_2019}. Recently, it has been shown that for GRBs, the value of $\beta$ can be positive or negative~\citep{Sahu_2020}. $\beta >0$ implies that the seed photons are in the self-Compton regime. $\beta < 0$ locates seed photons in the synchrotron regime.
It was previously shown that the VHE spectra of the GRB 190114C and GRB 190829A are due to the interaction of the high energy protons with the low energy tail region of the background synchrotron self Compton (SSC) photons in the jet with $\beta >0$~\citep{Sahu:2020dsg, Sahu:2022qaw}. Also shown there was that the VHE spectrum of GRB 180720B is from the interaction of high energy protons with the synchrotron seed photons in the jet environment with $\beta < 0$~\citep{Sahu:2020dsg}. This negative value of $\beta$ corresponds to the falling part of the synchrotron spectrum.

\section{Results}
LHAASO, with its two detectors WCDA and KM2A, detected $\ge 5000$ photons above 500 GeV from the GRB 221009A within $T\sim 2000$ s of the prompt emission.
The number of photons $N_{\gamma}$ detected at a time interval $T$ by any of these detectors at zenith angle $\theta$ and effective area $A(E_{\gamma},\theta)$ is~\citep{Zhao:2022wjg}
\beq
N_{\gamma}=T \int_{0.5\, TeV} \frac{dN_{\gamma}}{dE_{\gamma}} A(E_{\gamma},\theta)\, e^{-\tau_{\gamma}(E_{\gamma})} dE_{\gamma},
\label{ngamma}
\eeq
where the differential photon spectrum can be written as 
\beq
\frac{dN_{\gamma}}{dE_{\gamma}}=F_0\, E^{-\delta+1}_{\gamma,TeV}\, TeV^{-2}.
\eeq
The source was observed at a zenith angle of $30^{\circ}\lesssim \theta \lesssim 35^{\circ}$ that we adopt in Eq.(\ref{ngamma}). Taking into account the areas of LHAASO-WCDA and LHAASO-KM2A~\citep{LHAASO:2019qtb}, we evaluate the integral in Eq.(\ref{ngamma}) for $\delta=2.5,\,1.7$ and $1.2$. For the present analysis, we consider the EBL model of~\cite{Franceschini:2008tp}. We assume that these two detectors observe photons above 500 GeV in the range $5000 \le N_{\gamma}\le 6500$. By fixing the value of $N_{\gamma}$, we calculate the value of $F_0$ which is then used to calculate the VHE photon flux and the integrated flux $F^{int}_{\gamma}$ in the energy range $100\, \mathrm{GeV} \le E_{\gamma} \le 18\,\mathrm{TeV}$.

\begin{figure}
\centering
  \includegraphics[width=.85\linewidth]{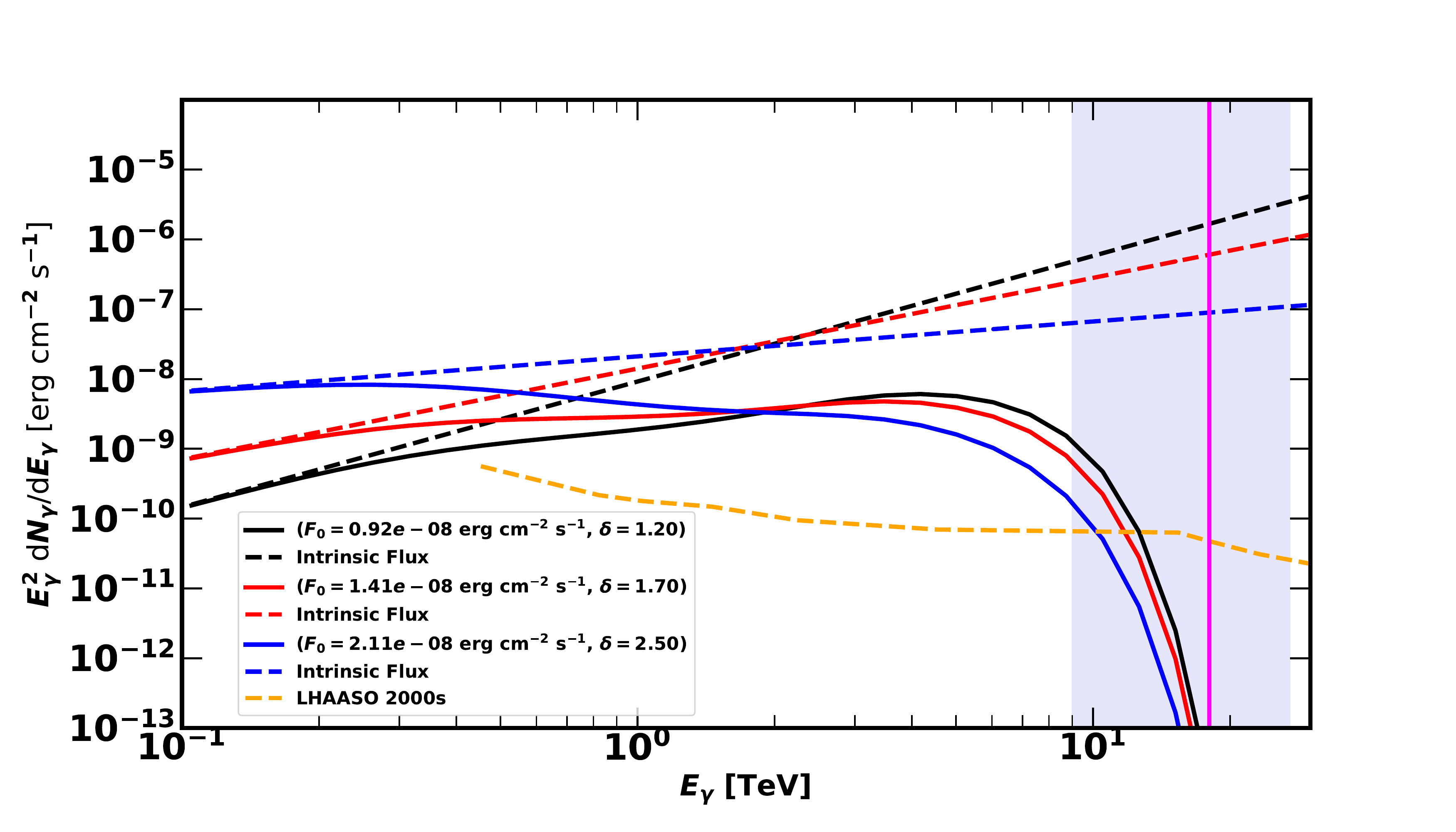}
\caption{Using the effective area of the detector LHAASO-WCDA the VHE Spectrum for GRB 221009A is given for different values of the spectral index $\delta$, by fixing $N_{\gamma}=5500$.
The intrinsic flux for each $\delta$ also shown. The LHAASO sensitivity curve for with 2000 s exposure is also shown. The vertical line corresponds to 18 TeV photon energy. The shaded region is $\pm 50\%$ relative energy resolution of LHAASO-WCDA for $E_{\gamma}\simeq 18$ TeV.
}
\label{fig:figure1}
\end{figure}

In Fig.~\ref{fig:figure1} we have shown the predicted spectra for $\delta=2.5, 1.7$ and $1.2$  by taking into account the effective area of the LHASSO-WCDA detector and fixing $N_{\gamma}=5500$. The relative energy resolution of LHAASO-WCDA is $\simeq 50\%$ at energies around 18 TeV (Fig. 26 of Chapter 1 of \cite{LHAASO:2019qtb}). For $\delta=2.5$ the flux starts from a maximum value of $F_{\gamma}\sim 10^{-8}\,\mathrm{erg\, cm^{-2}\, s^{-1}}$ at $E_{\gamma}=100$ GeV, and decreases slowly up to $\sim 4$ TeV. Beyond $\sim 4$ TeV it falls faster due to the EBL effect. The spectrum intersects with the sensitivity curve of LHAASO with 2000 s exposure at $E_{cut}=9.94$ TeV which is at the lower edge of the energy resolution (at 9 TeV). The $\delta=2.5$ value implies $\beta=0.5$ with the intrinsic flux $F_{\mathrm {in}}\propto E^{0.5}_{\gamma,TeV}$. This corresponds to seed photons in the lower tail region of the SSC spectrum in the GRB jet. The accelerated high energy protons in the jet interact with these seed photons to produce VHE gamma-rays, a situation very similar to the VHE flaring of HBLs. 

We repeat the calculation for $\delta=1.7$ which corresponds to $\beta=-0.3$. As discussed previously, negative value of the seed photon spectral index $\beta$ corresponds to photons in the descending part of the synchrotron spectrum towards higher $\epsilon_{\gamma}$ values and $\Phi_{\gamma}\propto \epsilon^{-0.3}_{\gamma}$. Thus, in this case, the high energy protons interact with the seed photons in the synchrotron regime of the external forward shock region to produce gamma-rays. The spectrum starts with $F_{\gamma}\sim 10^{-9} \mathrm{erg\, cm^{-2}\, s^{-1}}$ at $E_{\gamma}=100$ GeV and increases very slowly up to $\sim 4$ TeV and then falls faster as the exponentially decaying term from the EBL dominates. The curve intersects with the LHAASO sensitivity curve at $E_{cut}=11.53$ TeV. The intrinsic flux increases as  
$F_{\mathrm {in}}\propto E^{1.3}_{\gamma,TeV}$.

Finally, we consider a smaller value of $\delta=1.2$ which is shown in Fig.~\ref{fig:figure1}. This value of $\delta$ gives $\beta=-0.8$. In the photohadronic context, this corresponds to $\Phi_{\gamma}\propto \epsilon^{-0.8}_{\gamma}$ which is the descending part of the synchrotron spectrum towards higher $\epsilon_{\gamma}$ values like the ones for $\delta=1.7$. However, in this case, the seed synchrotron spectrum in the external forward shock region falls faster than the one for $\delta=1.7$. The spectrum increases and reaches a maximum flux at $E_{\gamma}\sim 4.5$ TeV, and then decreases exponentially for large values of $E_{\gamma}$ intersecting the LHAASO curve at $E_{cut} = 12.44$ TeV. The intrinsic flux in this case behaves like $E^{1.8}_{\gamma,TeV}$.

\begin{figure}
\centering
  \includegraphics[width=.85\linewidth]{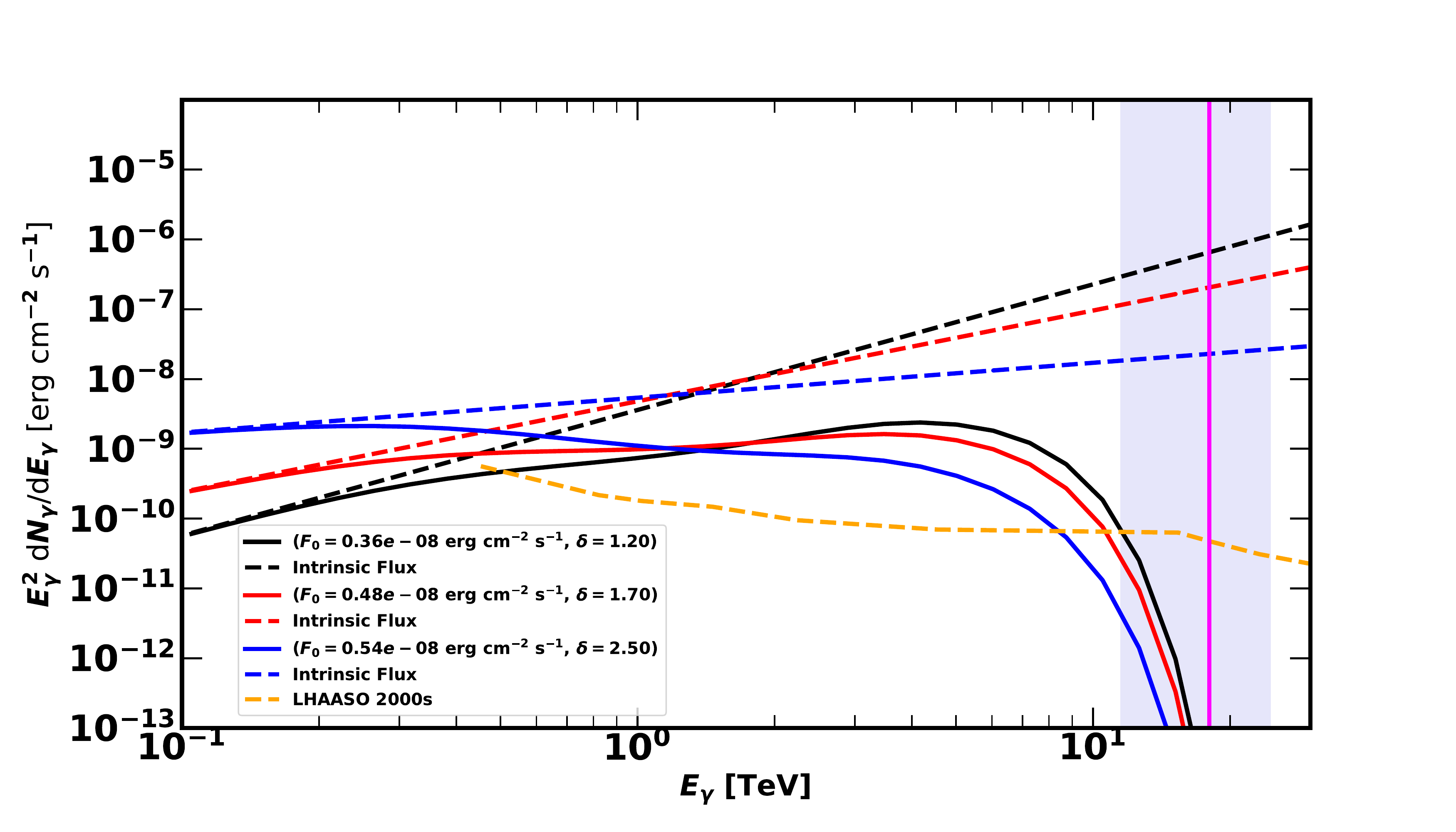}
\caption{This is same as Fig. \ref{fig:figure1} but using the detector area of LHAASO-KM2A and the shaded region here is $\pm 36\%$ relative energy resolution of LHAASO-KM2A for $E_{\gamma}\simeq 18$ TeV.}
\label{fig:figure2}
\end{figure}

\begin{table}
\centering
\caption{Using the LHAASO-WCDA ($30^{\circ}\le \theta \le 45^{\circ}$) and LHAASO-KM2A effective detector areas and different values of $\delta$ and number of events, $N_{\gamma}$, we have calculated the flux normalization factor $F_0$ in units of $10^{-8} \mathrm{erg\, cm^{-2}\, s^{-1}}$, the integrated flux $F^{int}_{\gamma}$ in units of $10^{-8} \mathrm{erg\, cm^{-2}\, s^{-1}}$, in the energy range 100 GeV to 18 TeV along with the corresponding luminosity $L_{\gamma, 48}$ in units of $10^{48}\, \mathrm{erg\, s^{-1}}$. $E_{cut}$ is the value of $E_{\gamma}$ in TeV unit where it intersects with the LHAASO sensitivity curve with 2000 s exposure time. The bracketed values are the results using the LHAASO-KM2A detector area. 
  }
\begin{tabular*}{\columnwidth}{@{\extracolsep{\fill}}llllll@{}}
\hline
$\delta$ & $N_{\gamma}$ & $F_0$ & $F^{int}_{\gamma}$ & $L_{\gamma,48}$ & $E_{cut}$ \\
\hline 
 2.5 & 5500 & 2.11 (0.54) & 2.49 (0.63) & 1.63 (0.41) &  9.94 (8.17)\\ 
     \, & 6500 & 2.50 (0.63) & 2.95 (0.75) &  1.92 (0.49) & 10.18 (8.41) \\
\hline 
1.7 & 5500 & 1.41 (0.48) & 1.22 (0.41) & 0.80 (0.27) & 11.53 (10.48) \\
      \, & 6500 & 1.67 (0.56) & 1.44 (0.49) & 0.94 (0.32) & 11.70 (10.63) \\ 
\hline 
1.2 & 5500 & 0.92 (0.36) & 1.07 (0.42) & 0.70 (0.27) & 12.44 (11.32) \\
      \, & 6500 & 1.08 (0.42) & 1.26 (0.50) & 0.83 (0.32) & 12.55 (11.50) \\ 
 \hline 
\end{tabular*}
\label{table1}
\end{table}

We repeat the calculation by using the effective area of LHAASO-KM2A for $\delta=2.5,\, 1.7,\, 1.2$ and $N_{\gamma}=5500$. The results are shown in Fig \ref{fig:figure2}. For 18 TeV photons the relative energy resolution of LHAASO-KM2A is $\simeq 36\%$ (Fig. 2, Chapter 1 of \cite{LHAASO:2019qtb}) which put the observed photon energy in the range $11.52$ TeV to $24.48$ TeV. For a given $\delta$, both LHAASO-WCDA and LHAASO-KM2A spectra the pattern is similar but the $E_{cut}$ value for LHAASO-KM2A is smaller than that of LHAASO-WCDA. Also, the $E_{cut}$ for LHAASO-KM2A is less than $11.52$ TeV which shows that LHAASO-KM2A may not be able to detect these photons.  

To account for the behavior of the VHE spectrum for different $N_{\gamma}$, we fix $N_{\gamma}=5500$ and $6500$ to calculate $F_0$, the integrated flux $F^{int}_{\gamma}$, the luminosity $L_{\gamma}$ and $E_{cut}$ values using the effective areas of the detectors LHAASO-WCDA and LHAASO-KM2A detectors. These are given in Table \ref{table1}. The results of LHAASO-KM2A are the bracketed values in the table. It can be seen that increasing $N_{\gamma}$ from $5500$ to $6500$ leads to increament in all quantities. This implies that by knowing $N_{\gamma}$ and the maximum value of $E_{\gamma}$ we can predict the VHE gamma-ray spectrum, provided the EBL contribution is well understood.

For a given value of $N_{\gamma}$, the $E_{cut}$ value increases and approaches $\sim 18$ TeV as $\delta$ decreases from 2.5 to 1.2. Moreover, by further decreasing $\delta$, one can reach $E_{cut}\sim 18$ TeV which corresponds to a very stiff synchrotron spectrum and may be problematic. Also, for a given $\delta$, by increasing $N_{\gamma}$, the $E_{cut}$ value increases. From our analysis we observed that, the LHAASO-WCDA is more likely to observe photons of energy $\sim 18$ TeV than the LHAASO-KM2A. From the dependence of $E_{cut}$ on $\delta$ we infer that the interaction of high energy protons with the descending part of the synchroton seed photon spectrum is more likely to produce $\sim 18$ TeV photons than the high energy protons interaction with the low energy tail region of the seed SSC photons in the GRB jet. 

\begin{figure}
\centering
  \includegraphics[width=.85\linewidth]{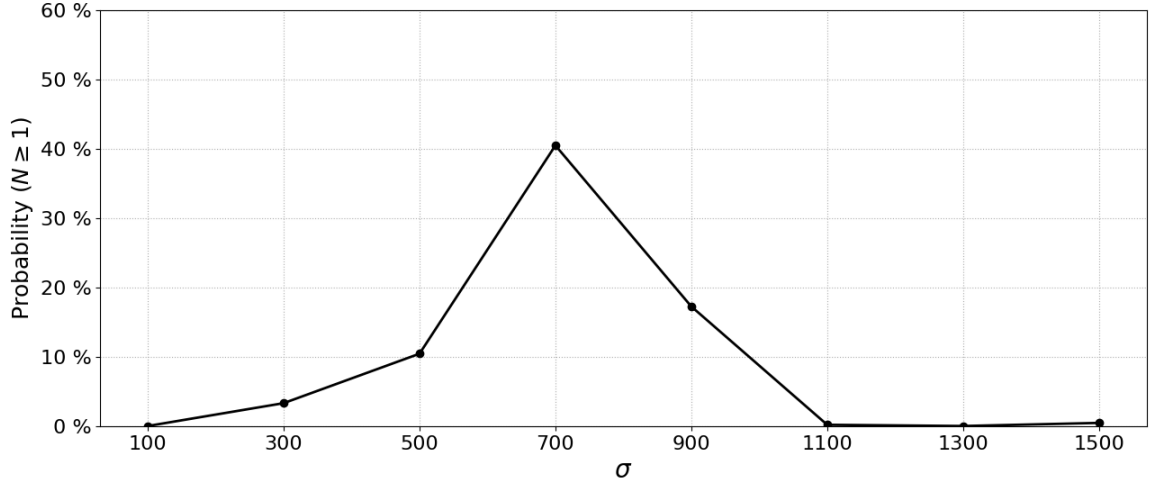}
\caption{The percentage of chance probability for $N_{\gamma}\ge 1$ for $E_{\gamma}=18$ TeV is plotted as a function of $\sigma$. 
}
\label{fig:figure3}
\end{figure}

Additionally, we calculate the chance probability of $N_{\gamma}\ge 1$ for $E_{\gamma} \sim 18$ TeV by taking the normalization constant $F_0$ as a variable in the range $10^{-10}\le F_0\, (\mathrm{erg\, cm^{-2}\, s^{-1}}) \le 2\times 10^{-7}$ for a fixed $\delta=1.2$. 
We fix the total number of observed photons above 500 GeV to be 5500 and $T=2000$ s. We assume that the noise in the data has a Gaussian distribution with an unknown standard deviation of $\sigma$ and variance $\sigma^2$~\citep{2010blda.book.....G}. Bayesian inference is implemented by using the Markov chain Monte Carlo (MCMC) method to estimate the posterior probability distribution function (PDF) as a function of $F_0$ for a given $\sigma$ value. This we have done for $\sigma$ in the range $100\le \sigma \le 1500$. Using the PDFs for different $\sigma$ values and $E_{\gamma}\sim 18$ TeV we evaluate 
\beq
N_{\gamma}(E_{\gamma})=T \int^{(1+\delta_E/2)E_{\gamma}}_{(1-\delta_E/2)E_{\gamma}} \frac{dN_{\gamma}}{dE'_{\gamma}} A(E'_{\gamma},\theta)\, e^{-\tau_{\gamma}(E'_{\gamma})} dE'_{\gamma},
\label{ng18tev}
\eeq
using the Monte Carlo simulation. Here $\delta_E=0.5$ is the value for the $50\%$ uncertainty in the energy around 18 TeV for the LHAASO-WCDA. For a given value of $\sigma$ we repeat the procedure for $10^6$ times. This way we obtain the percentage of chance probability of $N_{\gamma} \ge 1 = 10^{-4}\times \text{No\, of\, times\,} N_{\gamma}(18\, TeV) \ge 1$.
The results are plotted in Fig. \ref{fig:figure3}. It is observed that if the posterior PDF is symmetric 
around the mid-point, then the percentage of chance probability of detection of $N_{\gamma} \ge 1$ is very small. Similarly, for an asymmetric PDF with the weight factor leaning more towards smaller values of $F_0$, the percentage of chance probability is also very small.
However, for the asymmetric PDFs with the weight factor leaning more towards larger values of $F_0$, the percentage of chance probability is large with a maximum value of 40\%. 

\section{Conclusion}
In conclusion, the recent observation of $\sim 18$ TeV photons by LHAASO from GRB 221009A sheds doubts on the applicability of the well-known EBL models for photons of energy $> 10$ TeV at a redshift of $\gtrsim 0.151$ even though these EBL models work very well in explaining the VHE spectra of so many other TeV sources. This incompatibility has led towards new physics solutions. However, there is still a conventional way to delve into the problem, which we proposed here. We argue that high energy protons interacting with the synchrotron photon background in the GRB jet will be able to produce photons of energy close to 18 TeV. Assuming that the error in the data has a Gaussian distribution and using the area of LHAASO-WCDA, we obtain maximum 40\% chance probability of observing $N_{\gamma}\ge 1$ for $E_{\gamma}\sim 18$ TeV.
Our analysis shows that LHAASO-WCDA is more likely to observe photons of energy $\sim 18$ TeV than the LHAASO-KM2A. We anticipate that the publication of GRB 221009A results should be able to either confirm or rule out either most or all of the explanations discussed here. 



The authors are indebted to Miguel Enrique Iglesias Martínez and Jose Guerra Carmenate for helping in the Bayesian analysis. We are also thankful to the anonymous referee for making useful suggestions. S.S. is thankful to S. Salokya for reading the manuscript. The work of S.S. is partially supported by DGAPA-UNAM (Mexico) Projects No. IN103522. B. M-C and G. S-C would like to thank CONACyT (México) for partial support. Partial support from CSU-Long Beach is gratefully acknowledged. 


\bibliography{grbref}{}

\begin{thebibliography}{}
\expandafter\ifx\csname natexlab\endcsname\relax\def\natexlab#1{#1}\fi
\providecommand{\url}[1]{\href{#1}{#1}}
\providecommand{\dodoi}[1]{doi:~\href{http://doi.org/#1}{\nolinkurl{#1}}}
\providecommand{\doeprint}[1]{\href{http://ascl.net/#1}{\nolinkurl{http://ascl.net/#1}}}
\providecommand{\doarXiv}[1]{\href{https://arxiv.org/abs/#1}{\nolinkurl{https://arxiv.org/abs/#1}}}

\bibitem[{{Ackermann} {et~al.}(2012){Ackermann}, {Ajello}, {Allafort},
  {Schady}, {Baldini}, {Ballet}, {Barbiellini}, {Bastieri}, {Bellazzini},
  {Blandford}, {Bloom}, {Borgland}, {Bottacini}, {Bouvier}, {Bregeon},
  {Brigida}, {Bruel}, {Buehler}, {Buson}, {Caliandro}, {Cameron}, {Caraveo},
  {Cavazzuti}, {Cecchi}, {Charles}, {Chaves}, {Chekhtman}, {Cheung}, {Chiang},
  {Chiaro}, {Ciprini}, {Claus}, {Cohen-Tanugi}, {Conrad}, {Cutini},
  {D'Ammando}, {de Palma}, {Dermer}, {Digel}, {do Couto e Silva},
  {Dom{\'\i}nguez}, {Drell}, {Drlica-Wagner}, {Favuzzi}, {Fegan}, {Focke},
  {Franckowiak}, {Fukazawa}, {Funk}, {Fusco}, {Gargano}, {Gasparrini},
  {Gehrels}, {Germani}, {Giglietto}, {Giordano}, {Giroletti}, {Glanzman},
  {Godfrey}, {Grenier}, {Grove}, {Guiriec}, {Gustafsson}, {Hadasch},
  {Hayashida}, {Hays}, {Jackson}, {Jogler}, {Kataoka}, {Kn{\"o}dlseder},
  {Kuss}, {Lande}, {Larsson}, {Latronico}, {Longo}, {Loparco}, {Lovellette},
  {Lubrano}, {Mazziotta}, {McEnery}, {Mehault}, {Michelson}, {Mizuno}, {Monte},
  {Monzani}, {Morselli}, {Moskalenko}, {Murgia}, {Tramacere}, {Nuss},
  {Greiner}, {Ohno}, {Ohsugi}, {Omodei}, {Orienti}, {Orlando}, {Ormes},
  {Paneque}, {Perkins}, {Pesce-Rollins}, {Piron}, {Pivato}, {Porter},
  {Rain{\`o}}, {Rando}, {Razzano}, {Razzaque}, {Reimer}, {Reimer}, {Reyes},
  {Ritz}, {Rau}, {Romoli}, {Roth}, {S{\'a}nchez-Conde}, {Sanchez}, {Scargle},
  {Sgr{\`o}}, {Siskind}, {Spandre}, {Spinelli}, {Stawarz}, {Suson},
  {Takahashi}, {Tanaka}, {Thayer}, {Thompson}, {Tibaldo}, {Tinivella},
  {Torres}, {Tosti}, {Troja}, {Usher}, {Vandenbroucke}, {Vasileiou},
  {Vianello}, {Vitale}, {Waite}, {Winer}, {Wood}, \&
  {Wood}}]{2012Sci...338.1190A}
{Ackermann}, M., {Ajello}, M., {Allafort}, A., {et~al.} 2012, Science, 338,
  1190, \dodoi{10.1126/science.1227160}

\bibitem[{Alves~Batista(2022)}]{AlvesBatista:2022kpg}
Alves~Batista, R. 2022.
\newblock \doarXiv{2210.12855}

\bibitem[{Atteia(2022)}]{Circular32793}
Atteia, J.-L. 2022, GCN Circ. 32793

\bibitem[{{Baktash} {et~al.}(2022){Baktash}, {Horns}, \&
  {Meyer}}]{2022arXiv221007172B}
{Baktash}, A., {Horns}, D., \& {Meyer}, M. 2022, arXiv e-prints,
  arXiv:2210.07172.
\newblock \doarXiv{2210.07172}

\bibitem[{Bissaldi {et~al.}(2022)Bissaldi, Omodei, \& Kerr}]{Circular32637}
Bissaldi, E., Omodei, N., \& Kerr, M. 2022, GCN Circ. 32637

\bibitem[{Brdar \& Li(2022)}]{Brdar:2022rhc}
Brdar, V., \& Li, Y.-Y. 2022.
\newblock \doarXiv{2211.02028}

\bibitem[{Cao {et~al.}(2022)}]{LHAASO:2019qtb}
Cao, Z., {et~al.} 2022, Chin. Phys. C, 46, 035001.
\newblock \doarXiv{1905.02773}

\bibitem[{Cheung(2022)}]{Cheung:2022luv}
Cheung, K. 2022.
\newblock \doarXiv{2210.14178}

\bibitem[{Cortina(2005)}]{Cortina:2004qt}
Cortina, J. 2005, Astrophys. Space Sci., 297, 245,
  \dodoi{10.1007/s10509-005-7627-5}

\bibitem[{Das \& Razzaque(2022)}]{Das:2022gon}
Das, S., \& Razzaque, S. 2022.
\newblock \doarXiv{2210.13349}

\bibitem[{de~Ugarte~Postigo {et~al.}(2022)de~Ugarte~Postigo, Izzo, Pugliese,
  {et~al.}}]{Circular32648}
de~Ugarte~Postigo, A., Izzo, L., Pugliese, G., {et~al.} 2022, GCN Circ. 32648

\bibitem[{{Dermer} \& {Schlickeiser}(1993)}]{1993ApJ...416..458D}
{Dermer}, C.~D., \& {Schlickeiser}, R. 1993, \apj, 416, 458,
  \dodoi{10.1086/173251}

\bibitem[{Dominguez {et~al.}(2011)}]{Dominguez:2010bv}
Dominguez, A., {et~al.} 2011, MNRAS, 410, 2556,
  \dodoi{10.1111/j.1365-2966.2010.17631.x}

\bibitem[{{Dzhappuev} {et~al.}(2022){Dzhappuev}, {Afashokov}, {Dzaparova},
  {Dzhatdoev}, {Gorbacheva}, {Karpikov}, {Khadzhiev}, {Klimenko}, {Kudzhaev},
  {Kurenya}, {Lidvansky}, {Mikhailova}, {Petkov}, {Podlesnyi}, {Pozdnukhov},
  {Romanenko}, {Rubtsov}, {Troitsky}, {Unatlokov}, {Vaiman}, {Yanin}, \&
  {Zhuravleva}}]{2022ATel15669....1D}
{Dzhappuev}, D.~D., {Afashokov}, Y.~Z., {Dzaparova}, I.~M., {et~al.} 2022, The
  Astronomer's Telegram, 15669, 1

\bibitem[{{Finke} \& {Razzaque}(2022)}]{2022arXiv221011261F}
{Finke}, J.~D., \& {Razzaque}, S. 2022, arXiv e-prints, arXiv:2210.11261.
\newblock \doarXiv{2210.11261}

\bibitem[{{Fraija} {et~al.}(2022){Fraija}, {Gonzalez}, \& {HAWC
  Collaboration}}]{2022ATel15675....1F}
{Fraija}, N., {Gonzalez}, M., \& {HAWC Collaboration}. 2022, The Astronomer's
  Telegram, 15675, 1

\bibitem[{Franceschini {et~al.}(2008)Franceschini, Rodighiero, \&
  Vaccari}]{Franceschini:2008tp}
Franceschini, A., Rodighiero, G., \& Vaccari, M. 2008, A\&A, 487, 837,
  \dodoi{10.1051/0004-6361:200809691}

\bibitem[{Frederiks {et~al.}(2022)Frederiks, A.Lysenko, Ridnaia,
  {et~al.}}]{Circular32668}
Frederiks, D., A.Lysenko, Ridnaia, A., {et~al.} 2022, GCN Circ. 32668

\bibitem[{{Galanti} {et~al.}(2022){Galanti}, {Roncadelli}, \&
  {Tavecchio}}]{2022arXiv221005659G}
{Galanti}, G., {Roncadelli}, M., \& {Tavecchio}, F. 2022, arXiv e-prints,
  arXiv:2210.05659.
\newblock \doarXiv{2210.05659}

\bibitem[{Gehrels \& Razzaque(2013)}]{Gehrels:2013xd}
Gehrels, N., \& Razzaque, S. 2013, Front. Phys. (Beijing), 8, 661,
  \dodoi{10.1007/s11467-013-0282-3}

\bibitem[{Gotz {et~al.}(2022)Gotz, Mereghetti, Savchenko,
  {et~al.}}]{Circular32660}
Gotz, D., Mereghetti, S., Savchenko, V., {et~al.} 2022, GCN Circ. 32660

\bibitem[{{Gregory}(2010)}]{2010blda.book.....G}
{Gregory}, P. 2010, {Bayesian Logical Data Analysis for the Physical Sciences,
  Cambridge University Press, UK}

\bibitem[{Hauser \& Dwek(2001)}]{Hauser:2001xs}
Hauser, M.~G., \& Dwek, E. 2001, Ann. Rev. Astron. Astrophys., 39, 249,
  \dodoi{10.1146/annurev.astro.39.1.249}

\bibitem[{Hayes \& Gallagher(2022)}]{Hayes_2022}
Hayes, L.~A., \& Gallagher, P.~T. 2022, Research Notes of the {AAS}, 6, 222,
  \dodoi{10.3847/2515-5172/ac9d2f}

\bibitem[{Hinton(2004)}]{Hinton:2004eu}
Hinton, J.~A. 2004, New Astron. Rev., 48, 331,
  \dodoi{10.1016/j.newar.2003.12.004}

\bibitem[{Holder {et~al.}(2009)}]{Holder:2008ux}
Holder, J., {et~al.} 2009, AIP Conf. Proc., 1085, 657,
  \dodoi{10.1063/1.3076760}

\bibitem[{Huang {et~al.}(2022)Huang, Hu, Chen, {et~al.}}]{Circular32677}
Huang, Y., Hu, S., Chen, S., {et~al.} 2022, GCN Circ. 32677

\bibitem[{Kann \& Agui(2022)}]{Circular32762}
Kann, D.~A., \& Agui, J.~F. 2022, GCN Circ. 32762

\bibitem[{Krimm {et~al.}(2022)Krimm, Barthelmy, Dichiara,
  {et~al.}}]{Circular32688}
Krimm, H.~A., Barthelmy, S.~D., Dichiara, S., {et~al.} 2022, GCN Circ. 32688

\bibitem[{Lapshov {et~al.}(2022)Lapshov, Molkov, Mereminsky,
  {et~al.}}]{Circular32663}
Lapshov, I., Molkov, S., Mereminsky, I., {et~al.} 2022, GCN Circ. 32663

\bibitem[{{Li} \& {Ma}(2022)}]{2022arXiv221006338L}
{Li}, H., \& {Ma}, B.-Q. 2022, arXiv e-prints, arXiv:2210.06338.
\newblock \doarXiv{2210.06338}

\bibitem[{{Lin} \& {Yanagida}(2022)}]{2022arXiv221008841L}
{Lin}, W., \& {Yanagida}, T.~T. 2022, arXiv e-prints, arXiv:2210.08841.
\newblock \doarXiv{2210.08841}

\bibitem[{Meegan {et~al.}(2009)Meegan, Lichti, Bhat, Bissaldi, Briggs,
  Connaughton, Diehl, Fishman, Greiner, Hoover, van~der Horst, von Kienlin,
  Kippen, Kouveliotou, McBreen, Paciesas, Preece, Steinle, Wallace, Wilson, \&
  Wilson-Hodge}]{Meegan_2009}
Meegan, C., Lichti, G., Bhat, P.~N., {et~al.} 2009, ApJ., 702, 791,
  \dodoi{10.1088/0004-637X/702/1/791}

\bibitem[{{Mirabal}(2022)}]{2022arXiv221014243M}
{Mirabal}, N. 2022, arXiv e-prints, arXiv:2210.14243.
\newblock \doarXiv{2210.14243}

\bibitem[{Mitchell {et~al.}(2022)Mitchell, Phlips, Johnson,
  {et~al.}}]{Circular32746}
Mitchell, L.~J., Phlips, B.~F., Johnson, W., {et~al.} 2022, GCN Circ. 32746

\bibitem[{{Murase} {et~al.}(2022){Murase}, {Mukhopadhyay}, {Kheirandish},
  {Kimura}, \& {Fang}}]{2022arXiv221015625M}
{Murase}, K., {Mukhopadhyay}, M., {Kheirandish}, A., {Kimura}, S.~S., \&
  {Fang}, K. 2022, arXiv e-prints, arXiv:2210.15625.
\newblock \doarXiv{2210.15625}

\bibitem[{Nakagawa {et~al.}(2022)Nakagawa, Takahashi, Yamada, \&
  Yin}]{Nakagawa:2022wwm}
Nakagawa, S., Takahashi, F., Yamada, M., \& Yin, W. 2022.
\newblock \doarXiv{2210.10022}

\bibitem[{{Nemmen} {et~al.}(2012){Nemmen}, {Georganopoulos}, {Guiriec},
  {Meyer}, {Gehrels}, \& {Sambruna}}]{2012Sci...338.1445N}
{Nemmen}, R.~S., {Georganopoulos}, M., {Guiriec}, S., {et~al.} 2012, Science,
  338, 1445, \dodoi{10.1126/science.1227416}

\bibitem[{Piano {et~al.}(2022)Piano, Verrecchia, Bulgarelli,
  {et~al.}}]{Circular32657}
Piano, G., Verrecchia, F., Bulgarelli, A., {et~al.} 2022, GCN Circ. 32657

\bibitem[{Pillera {et~al.}(2022)Pillera, Bissaldi, Omodei,
  {et~al.}}]{Circular32658}
Pillera, R., Bissaldi, E., Omodei, N., {et~al.} 2022, GCN Circ. 32658

\bibitem[{Ripa {et~al.}(2022)Ripa, Pal, Werner, {et~al.}}]{Circular32685}
Ripa, J., Pal, A., Werner, N., {et~al.} 2022, GCN Circ. 32685

\bibitem[{Sahu(2019)}]{Sahu:2019lwj}
Sahu, S. 2019, Rev. Mex. Fis., 65, 307, \dodoi{10.31349/revmexfis.65.307}

\bibitem[{Sahu {et~al.}(2019)Sahu, Fort{\'{\i}}n, \& Nagataki}]{Sahu_2019}
Sahu, S., Fort{\'{\i}}n, C. E.~L., \& Nagataki, S. 2019, ApJ., 884, L17,
  \dodoi{10.3847/2041-8213/ab43c7}

\bibitem[{Sahu \& Fortín(2020)}]{Sahu_2020}
Sahu, S., \& Fortín, C. E.~L. 2020, ApJL., 895, L41,
  \dodoi{10.3847/2041-8213/ab93da}

\bibitem[{Sahu \& L\'opez~Fort\'\i{}n(2020)}]{Sahu:2020dsg}
Sahu, S., \& L\'opez~Fort\'\i{}n, C.~E. 2020, ApJL., 895, L41,
  \dodoi{10.3847/2041-8213/ab93da}

\bibitem[{Sahu {et~al.}(2020)Sahu, López Fortín, Iglesias Martínez,
  Nagataki, \& Fernández de Córdoba}]{10.1093/mnras/staa023}
Sahu, S., López Fortín, C.~E., Iglesias Martínez, M.~E., Nagataki, S., \&
  Fernández de Córdoba, P. 2020, MNRAS., 492, 2261,
  \dodoi{10.1093/mnras/staa023}

\bibitem[{Sahu {et~al.}(2022)Sahu, Polanco, \& Rajpoot}]{Sahu:2022qaw}
Sahu, S., Polanco, I. A.~V., \& Rajpoot, S. 2022, ApJ., 929, 70,
  \dodoi{10.3847/1538-4357/ac5cc6}

\bibitem[{S.Dichiara {et~al.}(2022)S.Dichiara, Gropp, Kennea,
  {et~al.}}]{Circular32632}
S.Dichiara, Gropp, J.~D., Kennea, J., {et~al.} 2022, GCN Circ. 32632

\bibitem[{{Stecker} {et~al.}(1992){Stecker}, {de Jager}, \&
  {Salamon}}]{1992ApJ...390L..49S}
{Stecker}, F.~W., {de Jager}, O.~C., \& {Salamon}, M.~H. 1992, \apjl, 390, L49,
  \dodoi{10.1086/186369}

\bibitem[{{Troitsky}(2022)}]{2022arXiv221009250T}
{Troitsky}, S.~V. 2022, arXiv e-prints, arXiv:2210.09250.
\newblock \doarXiv{2210.09250}

\bibitem[{Urry \& Padovani(1995)}]{Urry_1995}
Urry, C.~M., \& Padovani, P. 1995, Publications of the Astronomical Society of
  the Pacific, 107, 803, \dodoi{10.1086/133630}

\bibitem[{Ursi {et~al.}(2022)Ursi, Panebianco, Pittori,
  {et~al.}}]{Circular32650}
Ursi, A., Panebianco, G., Pittori, C., {et~al.} 2022, GCN Circ. 32650

\bibitem[{Veres {et~al.}(2022)Veres, Burns, Bissaldi, {et~al.}}]{Circular32636}
Veres, P., Burns, E., Bissaldi, E., {et~al.} 2022, GCN Circ. 32636

\bibitem[{{Wang} \& {Wei}(2011)}]{2011ApJ...726L...4W}
{Wang}, J., \& {Wei}, J.~Y. 2011, \apjl, 726, L4,
  \dodoi{10.1088/2041-8205/726/1/L4}

\bibitem[{Wu {et~al.}(2016)Wu, Zhang, Lei, Zou, Liang, \& Cao}]{Wu:2015opa}
Wu, Q., Zhang, B., Lei, W.-H., {et~al.} 2016, Mon. Not. Roy. Astron. Soc., 455,
  L1, \dodoi{10.1093/mnrasl/slv136}

\bibitem[{Wu {et~al.}(2011)Wu, Zou, Cao, Wang, \& Chen}]{Wu_2011}
Wu, Q., Zou, Y.-C., Cao, X., Wang, D.-X., \& Chen, L. 2011, ApJL., 740, L21,
  \dodoi{10.1088/2041-8205/740/1/L21}

\bibitem[{Xiao {et~al.}(2022)Xiao, Krucker, \& Daniel}]{Circular32661}
Xiao, H., Krucker, S., \& Daniel, R. 2022, GCN Circ. 32661

\bibitem[{Zhao {et~al.}(2022)Zhao, Zhou, \& Wang}]{Zhao:2022wjg}
Zhao, Z.-C., Zhou, Y., \& Wang, S. 2022.
\newblock \doarXiv{2210.10778}

\bibitem[{{Zhu} \& {Ma}(2022)}]{2022arXiv221011376Z}
{Zhu}, J., \& {Ma}, B.-Q. 2022, arXiv e-prints, arXiv:2210.11376.
\newblock \doarXiv{2210.11376}

\end{thebibliography}


\begin{references}
\end{references}
\bibliographystyle{aasjournal}

\end{document}